\documentclass{PoS}

\title{Optical Potential and Removal Energies in Lepton Nucleus Scattering }

\ShortTitle{Removal Energies}

\author{\speaker{Arie Bodek}\thanks{Supported by the US Department of Energy.}\\
        Department of Physics and Astronomy,  University of Rochester, Rochester, NY 14627 USA\\
        E-mail: \email{Bodek@pas.rochester.edu}}

\author{Tejin Cai\\
      Department of Physics and Astronomy, University of Rochester, Rochester, NY 14627 USA\\\
       E-mail: \email{tcai3@ur.rochester.edu}}

\abstract{We summarize some of the results presented in arXiv:1801.07975 [nucl-th]\cite{FSIpaper}(to be published in EPJC in 2018) on modeling 
electron and neutrino QE scattering  on a variety of nuclei within the impulse approximation.   We find that with three parameters we can describe the final state lepton energy for all of available electron QE data on Lithium, Carbon+Oxygen, Aluminum, Calcium+Argon, Iron and Lead+Gold. The first parameter, the removal energy $\epsilon^{P,N}$ is extracted from exclusive ee$^{\prime}$p spectral function data. The second parameter $V_{eff}$, which accounts for the interaction of  final state leptons and protons with the Coulomb potential of the nucleus,  is available from published comparisons of inclusive QE electron and positron cross section.  We extract the third parameter $U_{FSI}(\vec {q}_3^2)$, which accounts for the interaction of the final state nucleon with the optical potential of the spectator nucleus (FSI),  by fitting all available inclusive QE cross sections on nuclear targets. Here  $q_3$  is the three momentum transfer.  With these  three parameters we can model the energy of final state electrons and nucleons for all available electron QE scattering data.  At present the  uncertainty in the value of the removal energy parameters  is a the largest  source of systematic error in the extraction of the neutrino oscillation parameter $\Delta{m}^2$.  The use of the updated parameters in neutrino Monte Carlo generators reduces the systematic uncertainty in the combined removal energy (with FSI corrections) from $\pm$ 20 MeV to $\pm$ 5 MeV.  In this short contribution we only summarize the results for Carbon+Oxygen and Calcium+Argon.}
\FullConference{The 20th International Workshop on Neutrinos (NuFact2018)\\
		12-18 August 2018\\
		Blacksburg, Virginia}

\begin{document}
\begin{figure}[ht]
\begin{center}
\includegraphics[width=2.95in,height=2.5in]{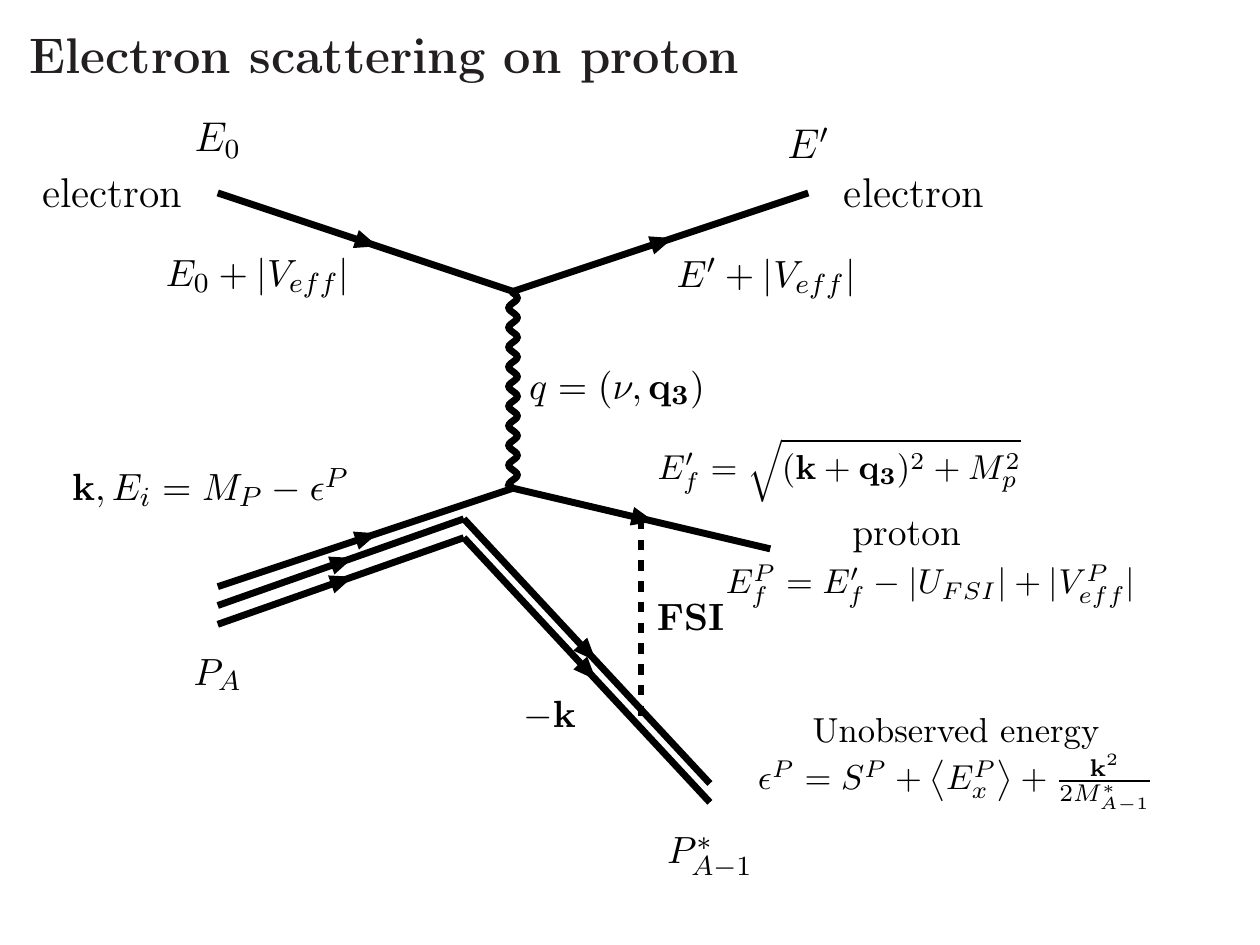}
\includegraphics[width=2.95in,height=2.5in]{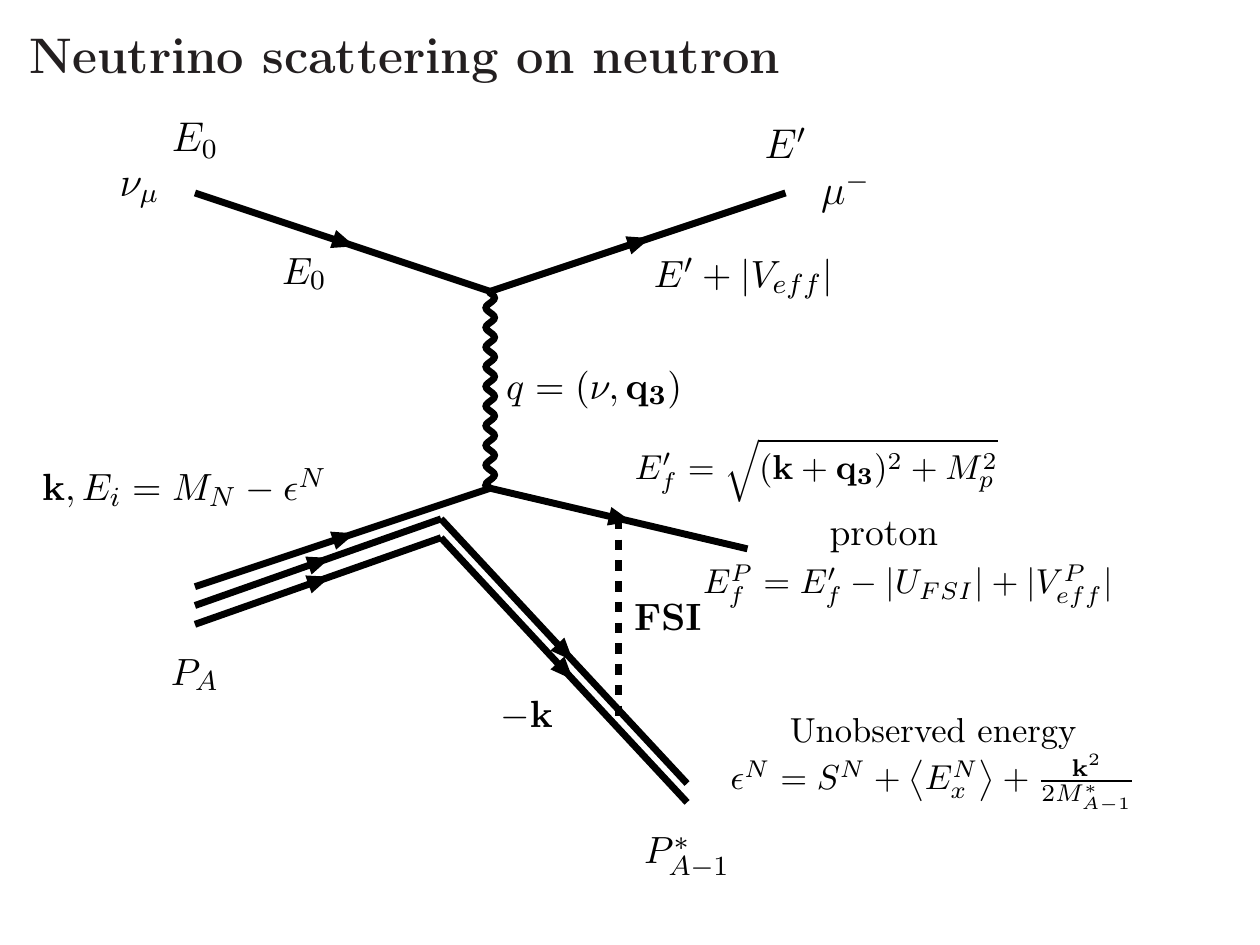}
   \vspace{-0.9cm}
\caption{
 \footnotesize\addtolength{\baselineskip}{-1\baselineskip} 
Electron (Left) and neutrino (right) QE  scattering from an off-shell bound proton. }
\label{Aoff-shell}
\end{center}
\end{figure}

We summarize some of the results\cite{FSIpaper} presented in arXiv:1801.07975 [nucl-th] (Removal Energies and Final State Interaction in Lepton Nucleus Scattering" to be published in EPJC in 2018) on modeling 
electron and neutrino quasielastic (QE) scattering  on a variety of nuclei within the impulse approximation.  Fig. \ref{Aoff-shell} shows the 
1p1h process for electron (left) and neutrino (right) QE  scattering from an off-shell bound nucleon of momentum $\vec {p_i}$=$\vec k$ in 
a nucleus of mass A.  Here, the nucleon is moving in the mean field (MF) of all  the other nucleons in the nucleus.  The on-shell recoil excited   $[A-1]^*$ spectator nucleus has a momentum  $\vec p_{ (A-1) *}=-\vec k$ and  a mean excitation energy  $\langle {E_x^{P,N}} \rangle$.
The off-shell energy of the interacting nucleon is  $E_i  =  M_A - \sqrt{ (M_{A-1}*) ^2+\vec k^2}=  M_A - \sqrt{ (M_{A-1}+{{E_x^{P,N}}})^2+\vec k^2} = M_P -\epsilon^{P,N} $, where  $\epsilon^P = S^{P,N} +\langle E_x^{P,N}  \rangle+\frac{\vec k^2}{2M^*_{A-1}}$.  Here $ S^{P,N}$ the separation energy 
 needed to separate a nucleon from the nucleus.

\begin{figure*}
\centering
\includegraphics[width=4.92cm,height=4.cm]{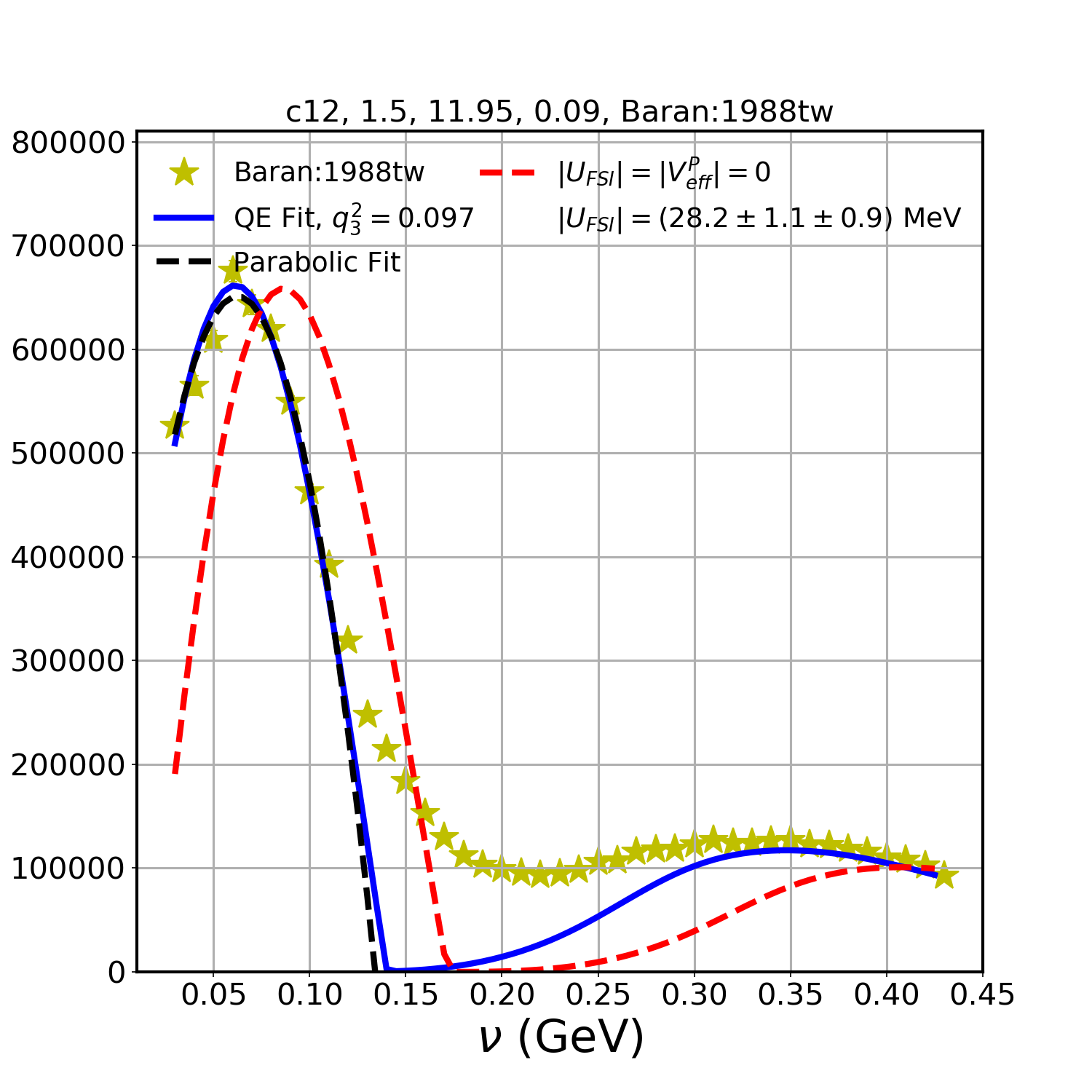}
\includegraphics[width=4.92cm,height=4.cm]{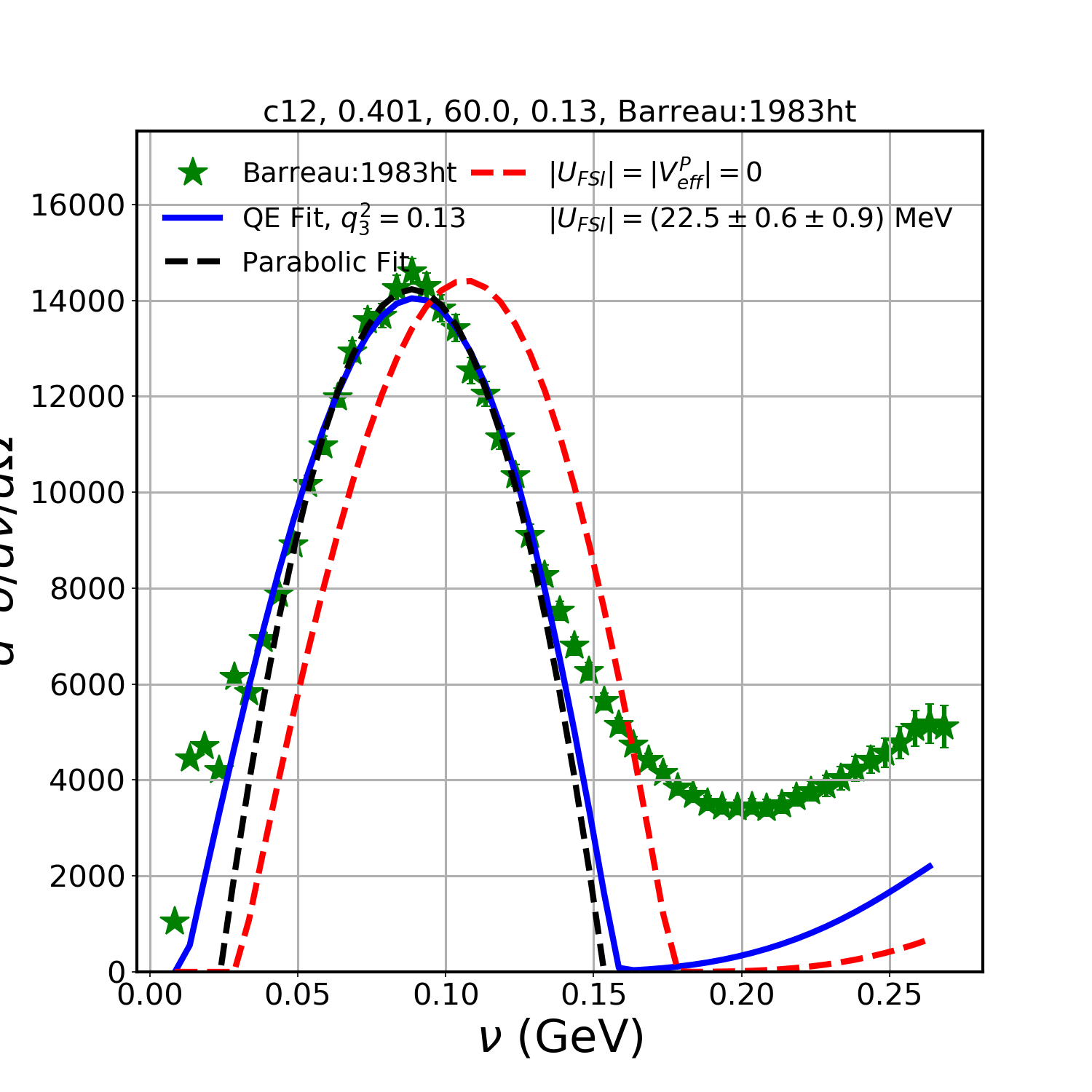}
\includegraphics[width=4.92cm,height=4.cm]{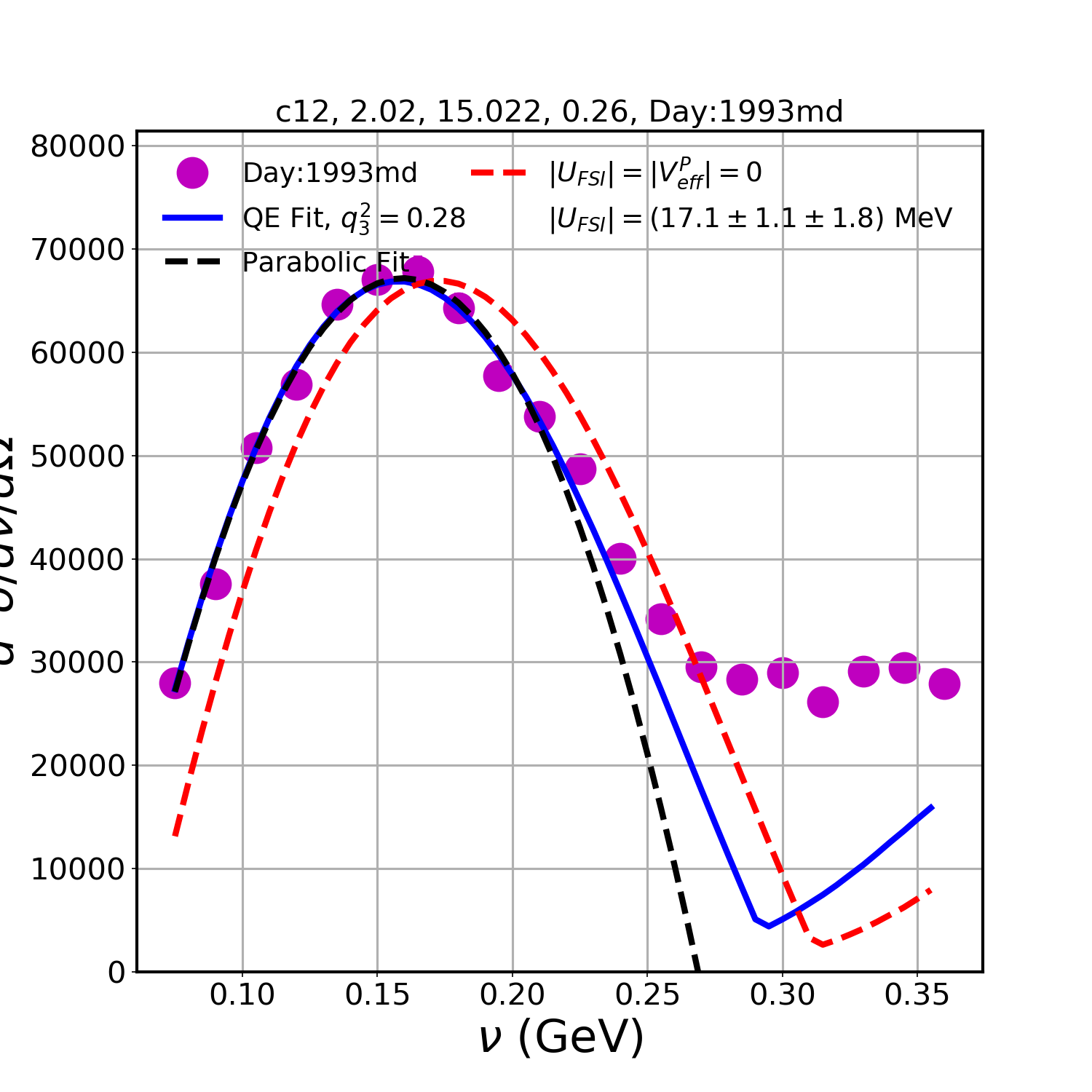}
   \vspace{-0.3cm}
\caption{
\footnotesize\addtolength{\baselineskip}{-1\baselineskip} 
Examples of fits for  three out of  33  carbon ($_{6}^{12}$C) QE differential cross sections. The solid blue curve is the RFG fit with the best value of $U_{FSI}$. The black dashed curve is the simple parabolic fit used to estimate the systematic error.  The red dashed curve is the RFG model  with  $U_{FSI}=V_{eff}=0$.
}
\label{C12_fits}
\end{figure*}

\begin{figure*} [ht]
\centering
  \includegraphics[width=15.cm,height=4.3cm]{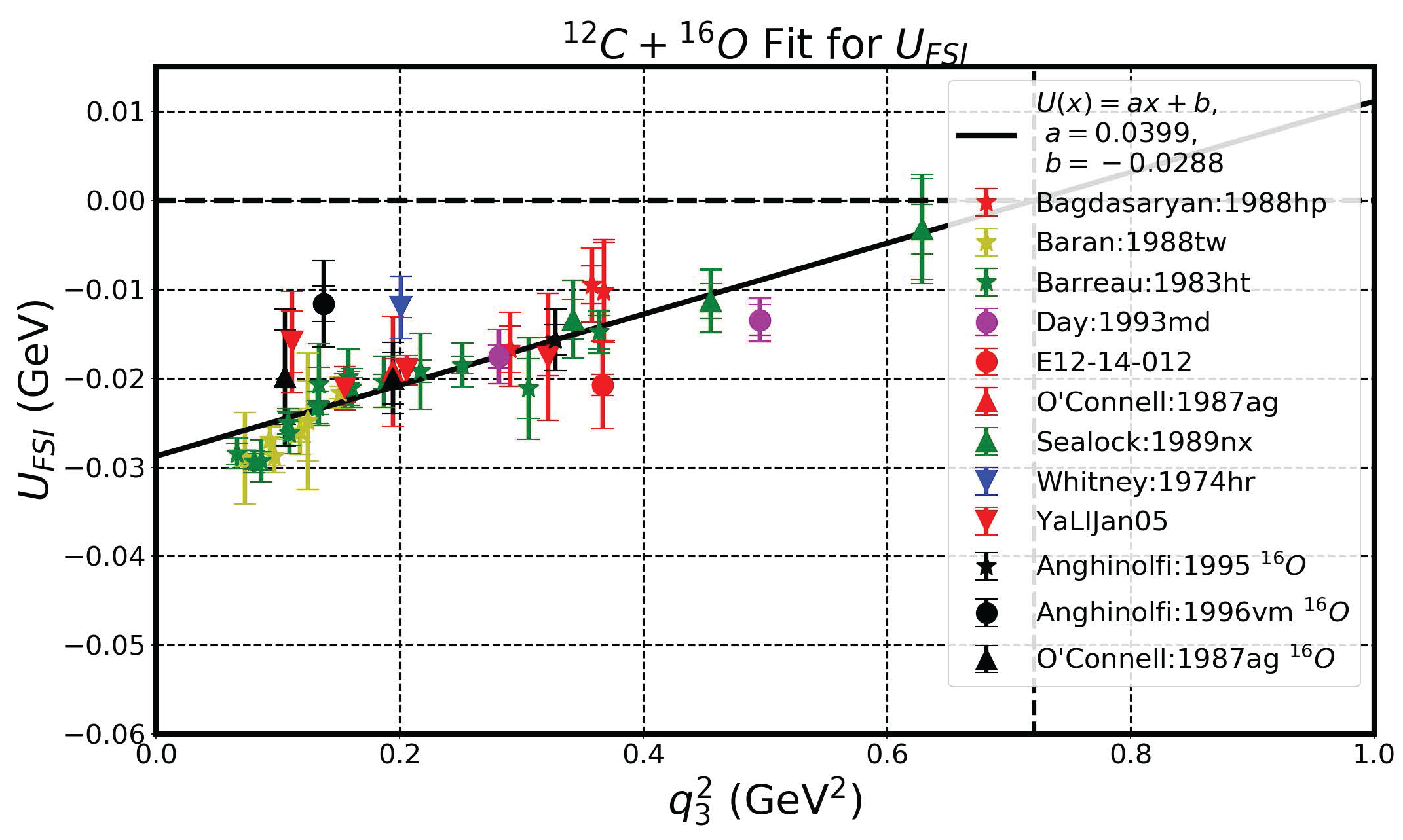}
       \includegraphics[width=15.cm,height=4.3cm]{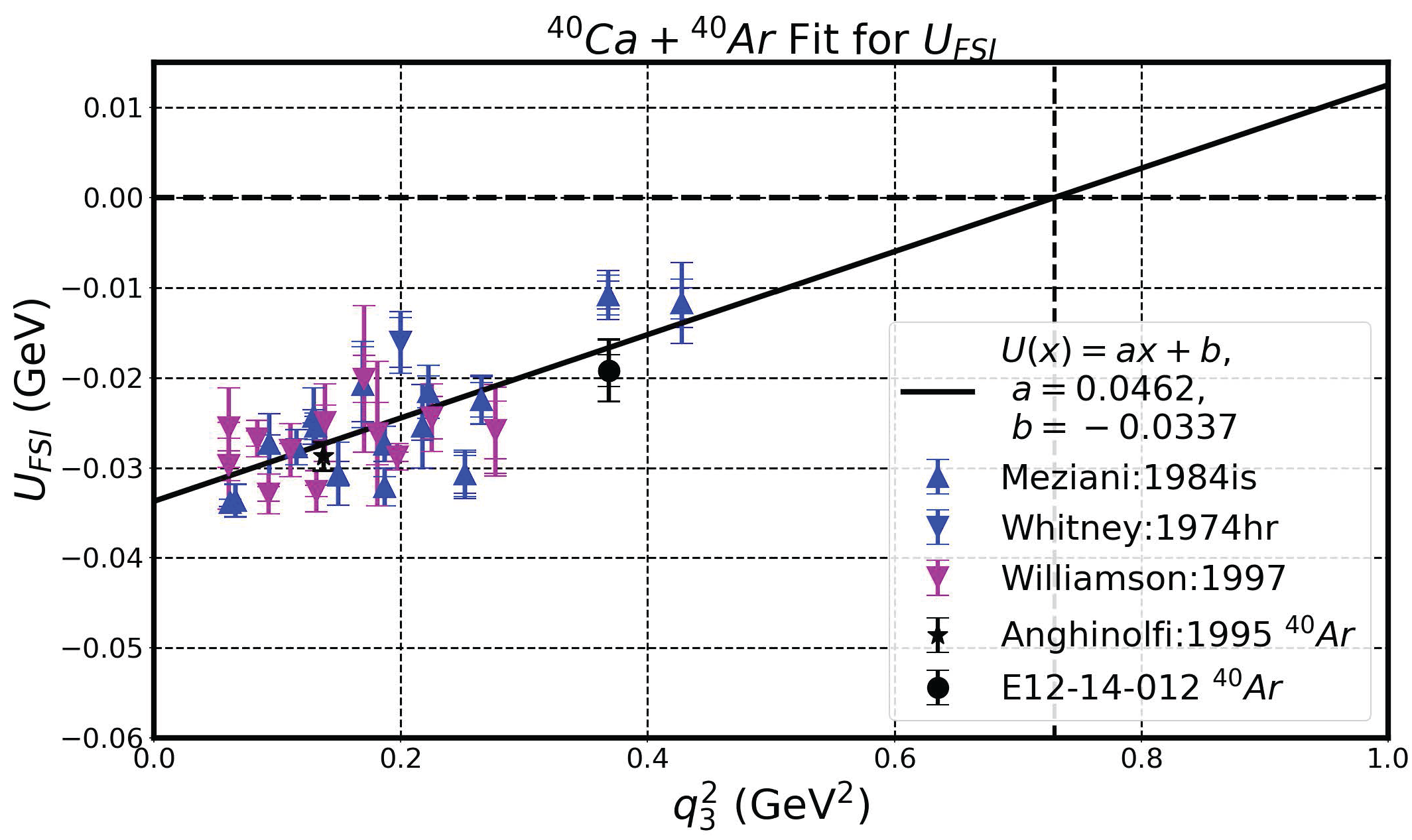}
       \vspace{-0.3cm}
\caption{
\footnotesize\addtolength{\baselineskip}{-1\baselineskip} 
Top plot: Extracted values of $U_{FSI}$ versus $\vec{q}_3^2$ for 33 Carbon ($_{6}^{12}$C) and four Oxygen  ($_{8}^{16}$O) spectra.
Bottom plot: Extracted values of $U_{FSI}$ versus $\vec {q}_3^2$ for 29 Calcium ($_{20}^{40}$Ca) and two Argon ( ($_{18}^{40}$Ar) spectra.}
  \label{C12vsq32}
\end{figure*}

Current neutrino MC generators (e.g. $\textsc{genie}$) do not include the effect of the excitation of the spectator nucleus, nor do they include the effects  of the interaction of the final state leptons and nucleons with the optical and Coulomb potential  of the nucleus. 
We extract the mean excitation energy  $\langle {E_x^P} \rangle$ from exclusive $ee^\prime P$ spectral function measurements. We include the effects  of the interaction of the final state leptons and protons with Coulomb field of the nucleus by using published parameters $V_{eff}$ obtained from a comparison of  electron and positron QE differential  cross sections\cite{veff} , and 
 model the effect of the interaction of final state nucleons with the optical potential of the nucleus  another parameter with  $|U_{FSI}(\vec q_3^2)|$. 
 We set energy  of a proton in the final state to  $E_f^P=\sqrt{(\vec {k}+\vec {q_3})^2+M_P^2} -|U_{FSI}(\vec q_3^2)|$+$|V_{eff}^P|$, and extract  
 $|U_{FSI}(\vec q_3^2)|$ from a comparison of the relativistic Fermi gas (RFG) model to measurements of inclusive QE e-A differential cross sections. 
\vspace{-0.2cm}
\vspace{-0.cm}
      The data samples  include:   four  $_{3}^{6}$Li spectra,     33  $_{6}^{12}$C spectra, five $_{8}^{16}$O spectra,    seven  $_{18}^{27}$Al spectra,   
     29  $_{20}^{40}$Ca spectra, two $_{18}^{40}$Ar spectra,   30 $_{26}^{56}$Fe  spectra,   23 $_{82}^{208}$Pb  spectra and one  $_{79}^{197}$Au  spectrum.  Most of the QE differential cross sections  are available on the  QE electron scattering archive\cite{archive}.  
 Figure \ref{C12_fits},  shows examples of three of 33 fits to QE  differential cross sections for $_{6}^{12}$C. The solid blue curve is the  RFG fit with the best value of $U_{FSI}$. The black dashed curve is a simple parabolic fit used to estimate the systematic error.  The red dashed curve is the RFG model  with  $U_{FSI}=V_{eff}=0$.
The extracted  values of  $U_{FSI}(\vec {q}_3^2)$ versus $\vec{q_3^2}$ for  $_{6}^{12}$C+$_{8}^{16}$O, and   $_{20}^{40}$Ca+$_{18}^{40}$Ar are shown in Figure \ref{C12vsq32}. 
 We fit the extracted values of $U_{FSI}(\vec {q}_3^2)$ versus $\vec{q_3^2}$ for $\vec{q_3^2}>0.1~GeV^2$ to a linear function. 
\vspace{-0.1cm}
The intercepts  at
$\vec{q_3^2}=0$ and the slopes of  $U_{FSI}(\vec q_3^2)$ are shown on the Figure.
     We only fit to the data in the top 1/3 of the QE distribution and extract  the best value of $U_{FSI} (\vec q_3^2)$.
  In the fit we let the normalization of the QE peak float to agree with data. 
The systematic error is estimated by using the difference between $\nu_{peak}^{parabola}$ and $\nu_{peak}^{rfg}$ as a systematic error in our
 extraction of $U_{FSI} (\vec q_3^2)$.  Here 
  $\vec q_3$ is evaluated at the peak of the QE distribution. With the  three parameters   
 we can model the energy of final state electrons and nucleons for all available electron QE scattering data. With the updated parameters the systematic uncertainty in the combined removal energy (with FSI corrections) in neutrino MC generators is reduced from $\pm$ 20 MeV to $\pm$ 5 MeV.
\vspace{-0.1cm}
\vspace{-0.7cm}
\end{document}